\documentclass[twocolumn]{aastex62}
\usepackage{array} 
\usepackage{amsmath}
\usepackage{wasysym}
\usepackage{mwe,tikz}
\usepackage[percent]{overpic}

\newenvironment{tightcenter}{%
	\setlength\topsep{0pt}
	\setlength\parskip{0pt}
	\begin{center}
}{%
 	\end{center}
}

\newcommand\chandra{{\it Chandra}}

\newcommand\ciao{CIAO}
\newcommand\sherpa{Sherpa}
\newcommand\xspec{XSPEC}

\newcommand\dof{\mathord{\rm dof}}

\newcommand\rhoc{\rho_{\rm S}}
\newcommand\rhol{\rho_{\rm L}}
\newcommand\zcboth{z_{{\rm S}, \pm}}
\newcommand\zcp{z_{{\rm S}, +}}
\newcommand\zcm{z_{{\rm S}, -}}
\newcommand\elobe{E_{\rm L}}
\newcommand\eshock{E_{\rm S}}
\newcommand\ecavity{E_{\rm C}}
\newcommand\rcavity{R_{\rm C}}

\newcommand\ie{i.e.}

\newcommand\thetajet{\theta_{\rm jet}}

\shorttitle{X-Ray Cavity Around Hotspot E in Cygnus\,A}

\shortauthors{Snios et al.}

\begin{document}

\title{The X-Ray Cavity Around Hotspot E in Cygnus\,A: Tunneled by a Deflected Jet}

\author{Bradford Snios} 
\affil{Harvard-Smithsonian Center for Astrophysics, 60 Garden Street, Cambridge, MA 02138, USA}
\author{Amalya C. Johnson} 
\affil{Harvard-Smithsonian Center for Astrophysics, 60 Garden Street, Cambridge, MA 02138, USA}
\affil{Columbia University, 538 W 120th Street, New York, NY 10027, USA}
\author{Paul E. J. Nulsen} 
\affil{Harvard-Smithsonian Center for Astrophysics, 60 Garden Street, Cambridge, MA 02138, USA}
\affil{ICRAR, University of Western Australia, 35 Stirling Hwy, Crawley, WA 6009, Australia}
\author{Ralph P. Kraft} 
\affil{Harvard-Smithsonian Center for Astrophysics, 60 Garden Street, Cambridge, MA 02138, USA}
\author{Martijn de Vries}
\affil{Department of Physics, Stanford University, 382 Via Pueblo Mall, Stanford, CA 94305, USA}
\author{Richard A. Perley}
\affil{National Radio Astronomy Observatory, P.O. Box 0, Socorro, NM 87801, USA}
\author{Lerato Sebokolodi}
\affil{National Radio Astronomy Observatory, P.O. Box 0, Socorro, NM 87801, USA}
\affil{Department of Physics and Electronics, Rhodes University, P.O. Box 94, Grahamstown, 6140, South Africa} 
\affil{South African Radio Astronomical Observatory, 2 Fir Street, Black River Park, Observatory, 7925, South Africa}
\author{Michael W. Wise}
\affil{Astronomical Institute ``Anton Pannekoek," University of Amsterdam, Postbus 94249, 1090 GE Amsterdam, The Netherlands}
\affil{SRON Netherlands Institute for Space Research, Sorbonnelaan 2, 3584CA Utrecht, The Netherlands}

\begin{abstract}
The powerful Fanaroff--Riley class II (FR\,II) radio galaxy Cygnus\,A exhibits primary and secondary hotspots in each lobe. A 2\,Msec \chandra{} X-ray image of Cygnus\,A has revealed an approximately circular hole, with a radius of 3.9\,kpc, centered on the primary hotspot in the eastern radio lobe, hotspot~E. We infer the distribution of X-ray emission on our line of sight from an X-ray surface brightness profile of the radio lobe adjacent to the hole and use it to argue that the hole is excavated from the radio lobe. The surface brightness profile of the hole implies a depth at least $1.7\pm0.3$ times greater than its projected width, requiring a minimum depth of $13.3\pm2.3$\,kpc. A similar hole observed in the 5\,GHz Very Large Array radio map reinforces the argument for a cavity lying within the lobe. We argue that the jet encounters the shock compressed intracluster medium at hotspot~E, passing through one or more shocks as it is deflected back into the radio lobe. The orientation of Cygnus\,A allows the outflow from hotspot~E to travel almost directly away from us, creating an elongated cavity, as observed. These results favor models for multiple hotspots in which an FR\,II jet is deflected at a primary hotspot, then travels onward to deposit the bulk of its power at a secondary hotspot, rather than the dentist drill model.
\end{abstract}


\section{Introduction}
\label{sect:intro}

Cygnus\,A, hereafter referred to as Cyg\,A, is widely regarded as the archetypal Fanaroff--Riley class II radio galaxy \citep[FR\,II;][]{Fanaroff1974}. At a redshift of $z=0.0561$ \citep{Owen1997} and an above average jet power of $10^{46}\rm\ ergs\ s^{-1}$ \citep{Godfrey2013, Snios2018b}, Cyg\,A has been extensively studied over a broad wavelength range \citep[e.g.,][]{Carilli1996}. The system is well known for its prominent radio lobes, which extend $\sim$\,65\arcsec{} from its central active galactic nucleus \citep[AGN;][]{Perley1984, Bartel1995}, as well as its cocoon shock that envelopes the system, as seen in the X-ray \citep{Carilli1994, Harris1994, Smith2002, Rafferty2006}. X-ray studies, in particular, provide a wealth of information on the energy transport from the jet to the surrounding intracluster medium \citep[ICM;][]{Snios2018b}. 

Distinctive brightness enhancements, or hotspots, are observed in Cyg\,A along the outer edges of its western and eastern lobes \citep{Carilli1991, Wright2004, Stawarz2007}. Each lobe has a primary hotspot (B and E) and a secondary hotspot (A and D), where the primary hotspots are defined to be more compact and less intense than the secondary hotspots \citep{Hargrave1974, Carilli1996}. The origin of the hotspots in Cyg\,A is a longstanding topic of debate \citep[e.g.,][]{Pyrzas2015}. One hotspot origin theory assumes a ``dentist drill" model in which the direction of the jet fluctuates over time, creating an actively fed hotspot (primary) and a remnant hotspot from previous jet/ICM interactions \citep[secondary;][]{Scheuer1982, Carilli1996, Steenbrugge2008}. While this mechanism explains the presence of multiple hotspots and the observed radio jet trajectory, the lifetime of the secondary hotspot is expected to be short once it is no longer fed by the jet \citep{Steenbrugge2008, Pyrzas2015}. It is therefore surprising that Cyg\,A possesses secondary hotspots given that their lifetimes are significantly less than the age of the system \citep{Pyrzas2015}, and that the secondary hotspots are notably more luminous than the primary hotspots \citep{Stawarz2007, Pyrzas2015}. Alternatively, the jet may deflect one or more times off the ICM before terminating in the outer lobe, creating a hotspot feature at each jet/ICM interaction region \citep{Williams1985, Cox1991}. This scenario explains the presence of multiple hotspots over longer timescales, assuming evidence can be shown of jetted flow between the hotspots. Here, we discuss X-ray signatures of interaction between the jet and plasma in Cyg\,A, finding evidence of a persistent outflow from primary hotspot E that favors the latter model.

This paper is one in a series on the analysis and interpretation of 2.0\,Msec of \chandra{} observations of Cyg\,A \citep{deVries2018, Duffy2018, Snios2018b}. Images made from the deep exposure show a deficit in surface brightness surrounding the primary hotspot in the eastern lobe (hotspot~E; Figure~\ref{fig:cyga}). The focus of this paper is to investigate this X-ray feature, learn what physical processes may explain its origin, and its implications for the standard hotspot model. The remainder of the paper is structured as follows. Section~\ref{sect:observation} outlines the details of the \chandra{} observations and the data reduction methods employed. Section~\ref{sect:sbp} describes surface brightness profiles extracted from the hole and surrounding lobe to establish the distribution of emission per unit volume along our line of sight. Spectra for the hotspots are extracted and fitted with emission models in Section~\ref{sect:flux} to measure their relative fluxes. The implications of our model fits are discussed in Section~\ref{sect:discuss}, and concluding remarks are given in Section~\ref{sect:conclusions}. 

For this work, we assume $H_{0} = 69.3\rm\ km\ s^{-1}\ Mpc^{-1}$, $\Omega_{M} = 0.288$, and $\Omega_{\Lambda} = 0.712$ \citep{Hinshaw2013}, which give an angular scale for Cyg\,A of $1.103\rm\ kpc\ arcsec^{-1}$ and an angular diameter distance of  227 Mpc at the redshift $z = 0.0561$. All uncertainties in the text are quoted at $1\sigma$ confidence intervals, unless otherwise specified. 

\section{Data Acquisition and Reduction} 
\label{sect:observation}

\begin{table}
	\caption{\textit{Chandra} Observations Used}
	\label{table:obs}
	\begin{tightcenter}
	{\footnotesize 
		\begin{tabular}{ c c c | c c c }
		\hline \hline
		ObsID & Date & $T_{\rm exp}$\tablenotemark{a} & ObsID & Date & $T_{\rm exp}$\tablenotemark{a} \\
		 &  & (ks) &  &  & (ks) \\
		\hline
		00360\tablenotemark{b} & 2000-05-21 & 34.3 & 17518 & 2016-07-16 & 49.4 \\
		01707\tablenotemark{b} & 2000-05-26 & 9.2 & 17519 & 2016-12-19 & 29.6 \\
		05830 & 2005-02-22 & 23.5 & 17520 & 2016-12-06 & 26.8 \\
		05831 & 2005-02-16 & 50.6 & 17521 & 2016-07-20 & 24.7 \\
		06225 & 2005-02-15 & 24.3 & 17522 & 2017-04-08 & 49.4 \\
		06226 & 2005-02-19 & 23.6 & 17523 & 2016-08-31 & 49.4 \\
		06228 & 2005-02-25 & 15.8 & 17524 & 2015-09-08 & 22.8 \\
		06229 & 2005-02-23 & 22.6 & 17525 & 2017-04-22 & 24.5 \\
		06250 & 2005-02-21 & 7.0 & 17526 & 2015-09-20 & 49.4 \\
		06252 & 2005-09-07 & 29.7 & 17527 & 2015-10-11 & 26.3 \\
		17133 & 2016-06-18 & 30.2 & 17528 & 2015-08-30 & 49.1 \\
		17134 & 2017-05-20 & 28.5 & 17529 & 2016-12-15 & 34.9 \\
		17135 & 2017-01-20 & 19.8 & 17530 & 2015-04-19 & 21.1 \\
		17136 & 2017-01-26 & 22.2 & 17650 & 2015-04-22 & 28.2 \\
		17137 & 2017-03-29 & 25.0 & 17710 & 2015-08-07 & 19.8 \\
		17138 & 2016-07-25 & 26.0 & 18441 & 2015-09-14 & 24.6 \\
		17139 & 2016-09-16 & 39.5 & 18641 & 2015-10-15 & 22.4 \\
		17140 & 2016-10-02 & 34.2 & 18682 & 2015-10-14 & 22.6 \\
		17141 & 2015-08-01 & 29.7 & 18683 & 2015-10-18 & 15.6 \\
		17142 & 2017-04-20 & 23.3 & 18688 & 2015-11-01 & 34.4 \\
		17143 & 2015-09-03 & 26.9 & 18871 & 2016-06-13 & 21.6 \\
		17144 & 2015-05-03 & 49.4 & 18886 & 2016-07-23 & 22.2 \\
		17507 & 2016-11-12 & 32.6 & 19888 & 2016-10-01 & 19.5 \\
		17508 & 2015-10-28 & 14.9 & 19956 & 2016-12-10 & 54.3 \\
		17509 & 2016-07-10 & 51.4 & 19989 & 2017-02-12 & 41.5 \\
		17510 & 2016-06-26 & 37.1 & 19996 & 2017-01-28 & 28.1 \\
		17511 & 2017-05-10 & 15.9 & 20043 & 2017-03-25 & 29.6 \\
		17512 & 2016-09-15 & 66.9 & 20044 & 2017-03-26 & 14.9 \\
		17513 & 2016-08-15 & 49.4 & 20048 & 2017-05-19 & 22.6 \\
		17514 & 2016-12-13 & 49.4 & 20059 & 2017-04-19 & 23.8 \\ 
		17515 & 2017-03-21 & 39.3 & 20063 & 2017-04-22 & 25.4 \\
		17516 & 2016-08-18 & 49.0 & 20077 & 2017-05-13 & 27.7 \\
		17517 & 2016-09-17 & 26.7 & 20079 & 2017-05-21 & 23.8 \\
		\hline
		\multicolumn{5}{r}{Total Exposure Time} & 2007.9 \\
		\hline
	\end{tabular}}
	\end{tightcenter}
	\tablenotemark{a}{Net exposure after background flare removal.\\
	\tablenotemark{b}{Observed with ACIS-S; all others
		observed with ACIS-I.}}
\end{table}

Cyg\,A is a bright X-ray source that has been repeatedly observed by \chandra{} over the telescope's lifetime \citep{Smith2002, Harris2006, Wilson2006, deVries2018, Duffy2018, Snios2018b}. We may therefore leverage the extensive archival database of Cyg\,A to generate a deep observation of the system. See Table~\ref{table:obs} for a complete list of observations used in this analysis. Since the point-spread function of \chandra{} is known to broaden away from the aimpoint, we selected observations where the aimpoint was within 1\arcmin{} of the central AGN. We note that ObsIDs 00360 and 01707 were taken with the telescope aimpoint centered on the S3 chip of the Advanced CCD Imaging Spectrometer (ACIS). Additionally, ObsID 00360 was performed in FAINT mode, while ObsID 01707 was performed in VFAINT mode. All remaining observations were completed with the ACIS-I array in FAINT mode. 

\begin{figure*}
  \begin{tightcenter}   
  \begin{overpic}[width=1.0\linewidth]{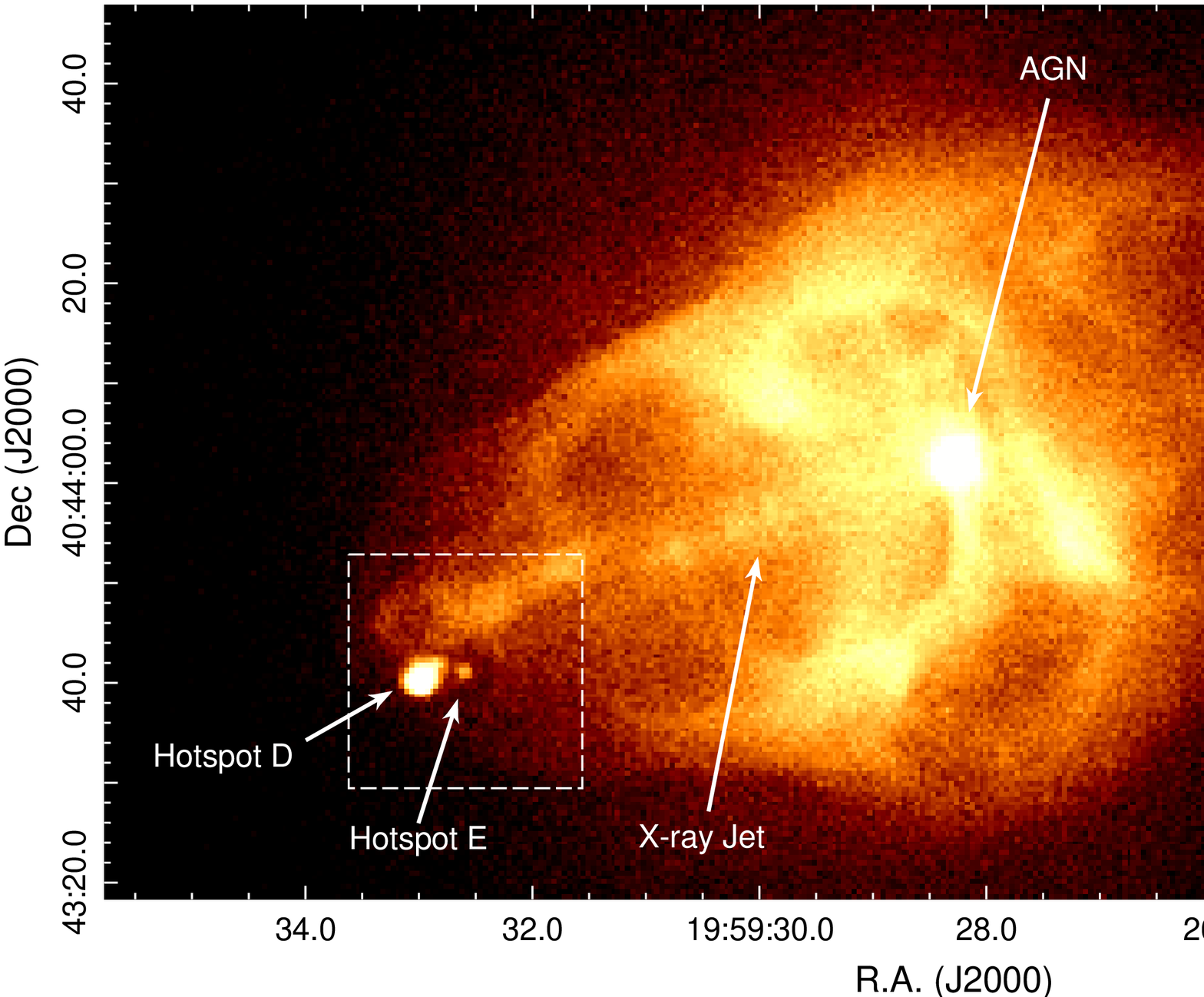}
     \put(7.5,30){\includegraphics[width=0.23\linewidth]{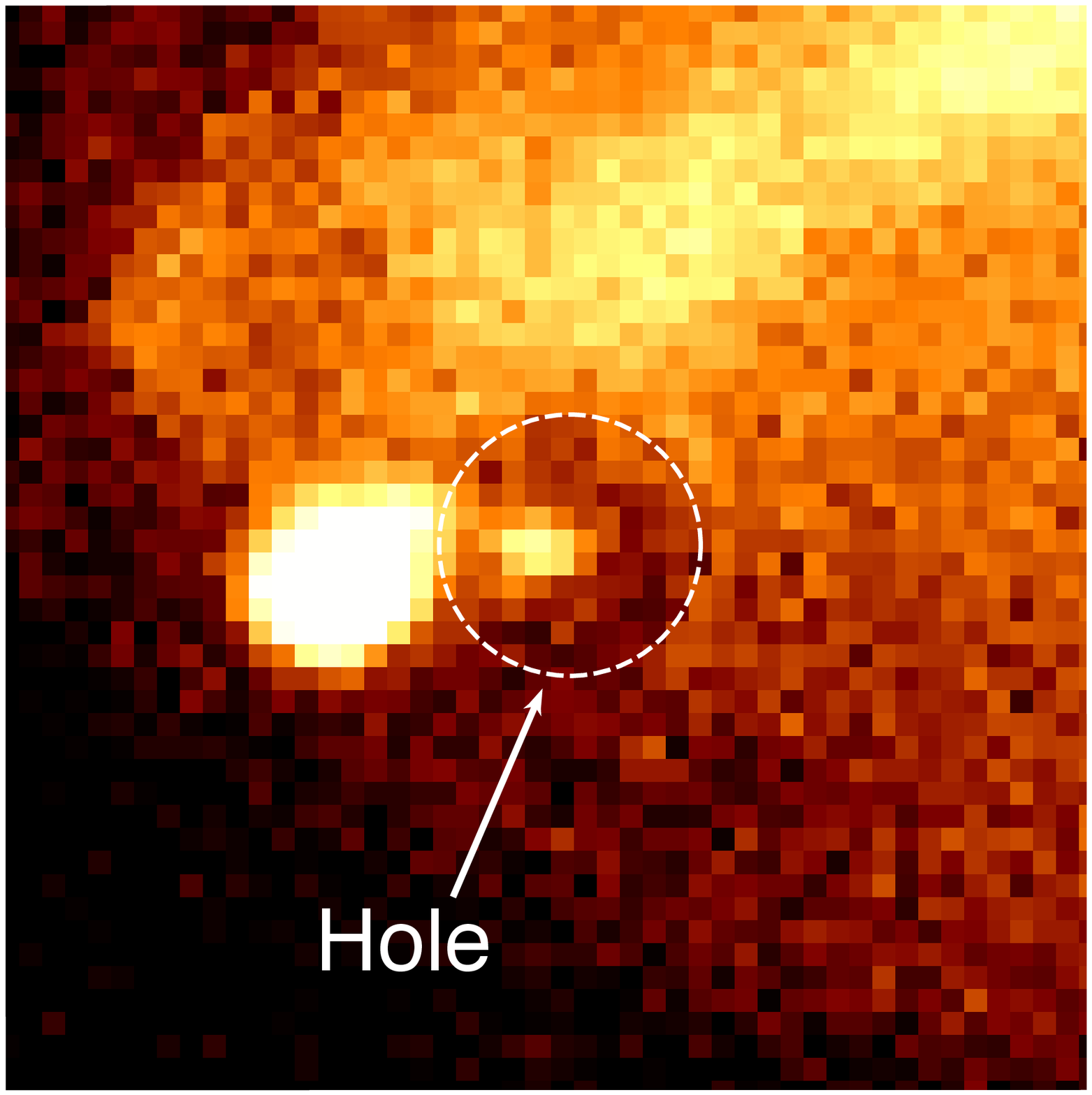}}  
  \end{overpic}
  \end{tightcenter}
\caption{A 0.5--7.0 keV, exposure-corrected \chandra{} image of Cygnus\,A. The observations listed in Table~\ref{table:obs} were co-added for the image, and the pixel size is 0.492\arcsec. A deficit in brightness is observed surrounding hotspot~E, which is enlarged and shown at higher contrast in the inset for additional clarity.}
\label{fig:cyga}
\end{figure*}

Correction for residual astrometric errors in the observations is required to minimize blurring in the final, merged image. We aligned the Cyg\,A dataset using the two-dimensional cross-correlation method described in \cite{Snios2018b}, including use of the same reference image (ObsID 05831) and fitted region. The resulting astrometric corrections were applied to each observation with the \ciao{} 4.10 task {\tt wcs\_update} \citep{Fruscione2006}, where the root mean square translations were $\Delta x_{\rm rms} = 0.82\arcsec{}$ and $\Delta y_{\rm rms} = 0.24\arcsec{}$. Each observation was additionally reprocessed using the \ciao{} routine {\tt deflare} to remove background flaring periods from the data, while the routine {\tt readout\_bkg} was used to estimate the distribution of ``out-of-time" events. The cleaned exposure time for each observation is provided in Table~\ref{table:obs}, resulting in a total exposure of 2.01\,Msec. 

Exposure maps for the dataset were created for the 0.5--7.0\,keV energy band assuming the absorbed thermal model ${\tt phabs} \times{\tt apec}$, with a temperature of 5.5\,keV and abundances of $0.66Z_{\odot}$ based on the solar ratios of \cite{Anders1989}. No background was subtracted from the images presented here, although local background is subtracted for the spectral analyses in Sections~\ref{sect:sbp} and \ref{sect:flux}. The merged, exposure-corrected image of Cyg\,A is shown in Figure~\ref{fig:cyga}. 

\section{Surface Brightness Profiles}
\label{sect:sbp}

Examination of Figure~\ref{fig:cyga} shows a deficit in brightness surrounding hotspot~E. The fact that this feature is centered on the hotspot is a strong indication that the two are physically associated. The X-ray deficit may indicate that outflow from the hotspot has carved a region out of the radio lobe, creating the observed surface brightness ``hole" in the image. An X-ray profile of the lobe is measured in Section~\ref{sect:lobe} and used to determine the emission per unit volume from the radio lobe. This result together with the distribution of X-ray emission around the hotspot is used to constrain X-ray emission from within the cavity in Section~\ref{sect:bubble}.

\subsection{Eastern Radio Lobe} 
\label{sect:lobe}

\begin{figure*}
  \begin{tightcenter}   
	\includegraphics[width=0.442\linewidth]{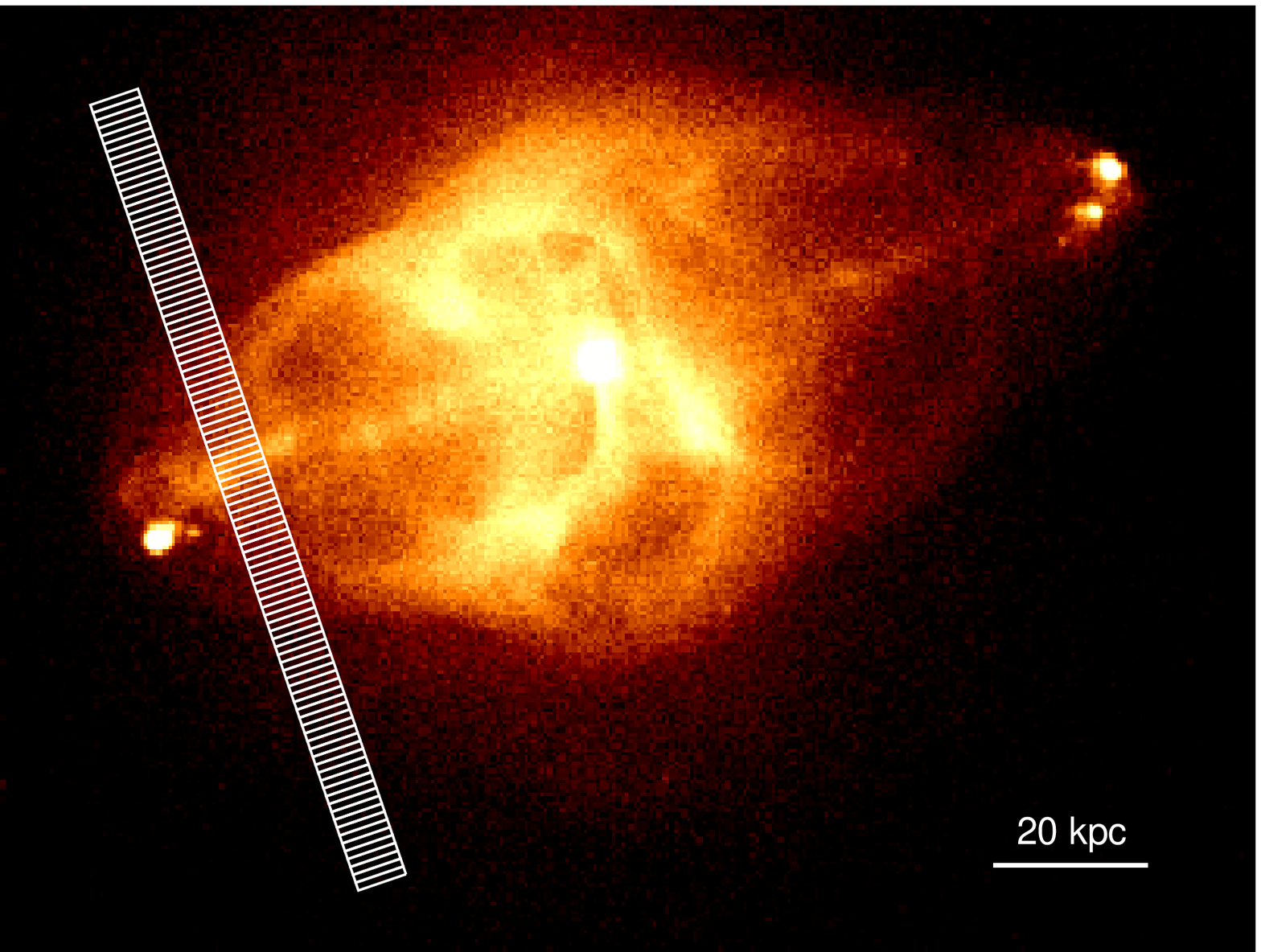}
	\includegraphics[width=0.55\linewidth]{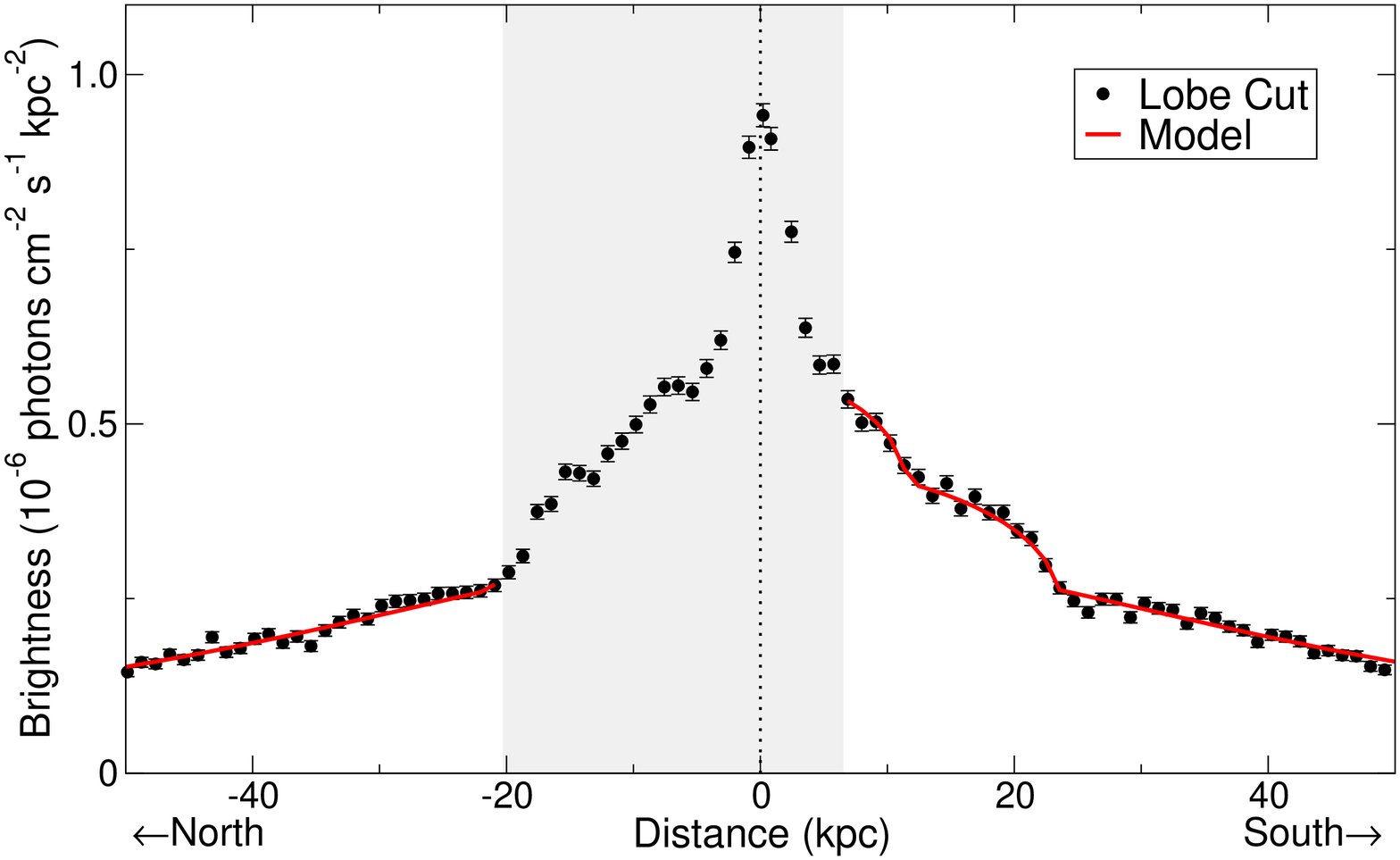}
  \end{tightcenter}
\caption{(Left) The rectangular region used for the surface brightness profile of the eastern radio lobe. (Right) The extracted surface brightness profile with the fitted model described in Section~\ref{sect:lobe}. The gray region was not fit with the model to avoid complications in modeling the X-ray jet and the asymmetric structure present in the northern half of the lobe.}
\label{fig:cut}
\end{figure*}

Emission projected onto the lobe regions is composed of thermal radiation from the hot intracluster gas, including a layer of shock compressed ICM immediately surrounding the lobe, and nonthermal, Inverse Compton radiation from the relativistic plasma within the lobe itself \citep{Wilson2006, deVries2018}. Assuming that the lobe and cluster emission is symmetric under rotation about the axis of the jets, the emission per unit volume distribution may be estimated by fitting a surface brightness profile perpendicular to the jets.

A rectangular region perpendicular to the jet axis of Cyg\,A was divided into 100 equal areas and used to measure the surface brightness profile (Figure~\ref{fig:cut}, left). The region was positioned to obtain the best possible estimate of the emission from the lobe close to the hole. The \ciao{} routine {\tt dmextract} was used to extract the surface brightness from the exposure-corrected image. A plot of the surface brightness profile is shown in Figure~\ref{fig:cut}, right. 

To determine the emission from within the radio lobe, a model was fitted to the surface brightness profile. Details of the model are discussed in Appendix~\ref{apen:model}, while the fitting method and best-fit parameters are discussed in Appendix~\ref{apen:fit}. For the purpose of the model, the cluster is divided into three regions. A beta model represents the emission from the ICM, a cylindrical shell of constant emission per unit volume represents the shock compressed ICM, and, within that, another cylinder of constant emission per unit volume representing emission from within the radio lobe. The model is projected onto the sky in order to simulate the surface brightness profile of the lobe. Emission from the X-ray jet, close to the radio axis, is clearly brighter than the remainder of the lobe \citep[de\,Vries et al., in preparation]{Steenbrugge2008}, so it was not included in the fit. Additionally, emission from the northern half of the radio lobe was excluded due to the observed asymmetry between the northern and southern sections, with the northern half being consistently brighter. These excluded regions are indicated by gray shading in Figure~\ref{fig:cut}, right. 

Despite its simplicity, the model fits the surface brightness profile well. Although the X-ray emission per unit volume is significantly greater within the lobe than in the shocked ICM, we see no evidence for significant gradients in the emission per unit volume within these regions. The inclination of the axis of the radio lobe to our line of sight was varied to study its effect on estimates of emission per unit volume, and the best-fit parameters varied by $< 3\sigma$ for angles in the range \mbox{45--90$^{\circ}$}. The emission per unit volume from the lobe and shocked ICM are both maximized for an inclination of $90^\circ$ and we use these values in the following discussion, minimizing estimates for the depth of the cavity. The best-fit emission per unit volume from the sheath of shock compressed gas is $6.82\substack{+0.24\\-0.33}\,\times\,10^{-9} \rm\ photons\ cm^{-2}\ arcsec^{-3}\ s^{-1}$. From the radio lobe, it is $1.42\substack{+0.07\\-0.06}\,\times\,10^{-8} \rm\ photons\ cm^{-2}\ arcsec^{-3}\ s^{-1}$. Best-fit parameters for line of sight inclination angles of $90^{\circ}$ and $55^{\circ}$ \citep{Vestergaard1993}, are provided in Appendix~\ref{apen:fit}.

\subsection{Hotspot~E Cavity} 
\label{sect:bubble}

\begin{figure*}
  \begin{tightcenter}   
	\includegraphics[width=0.50\linewidth]{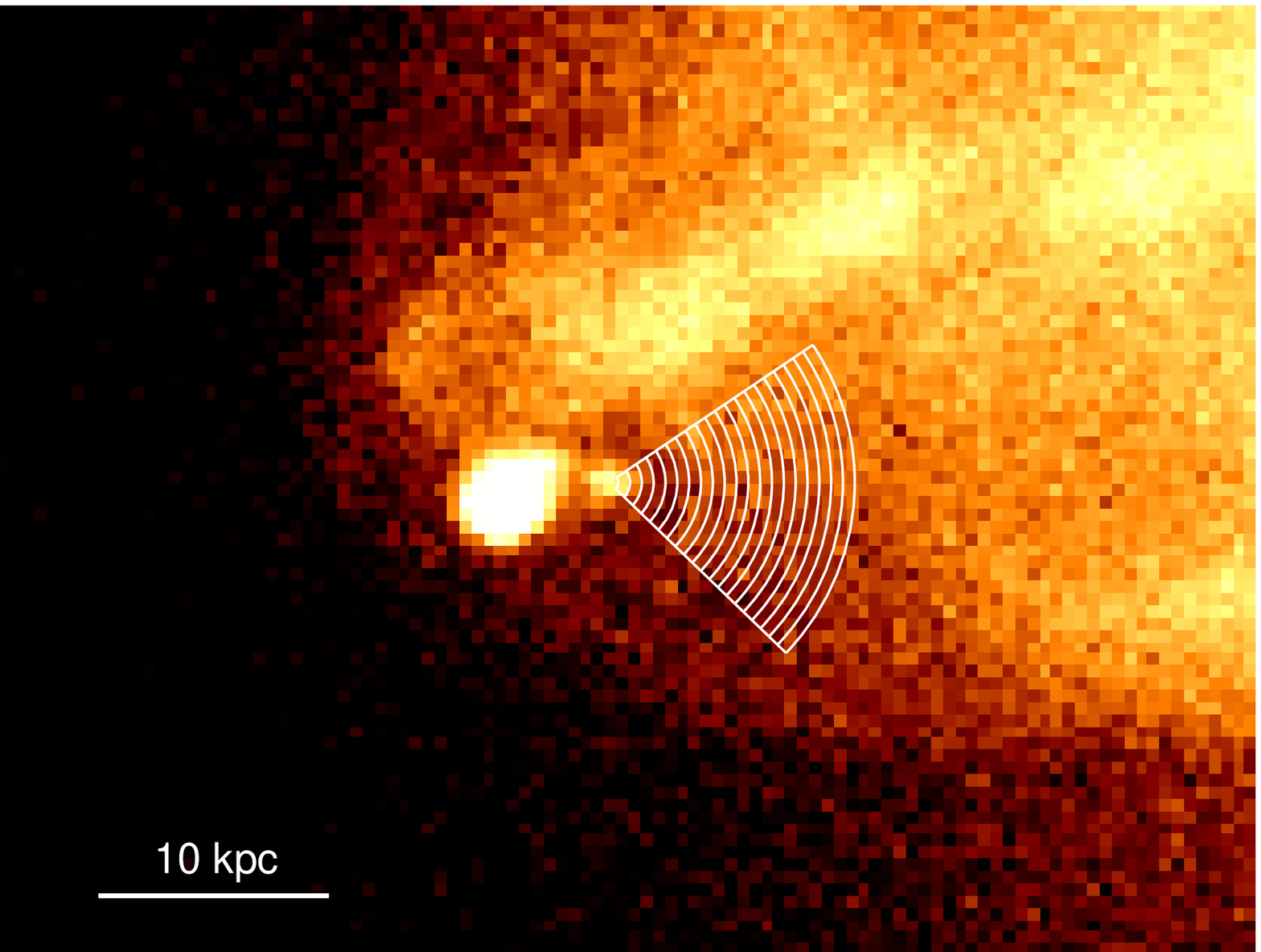}
	\includegraphics[width=0.49\linewidth]{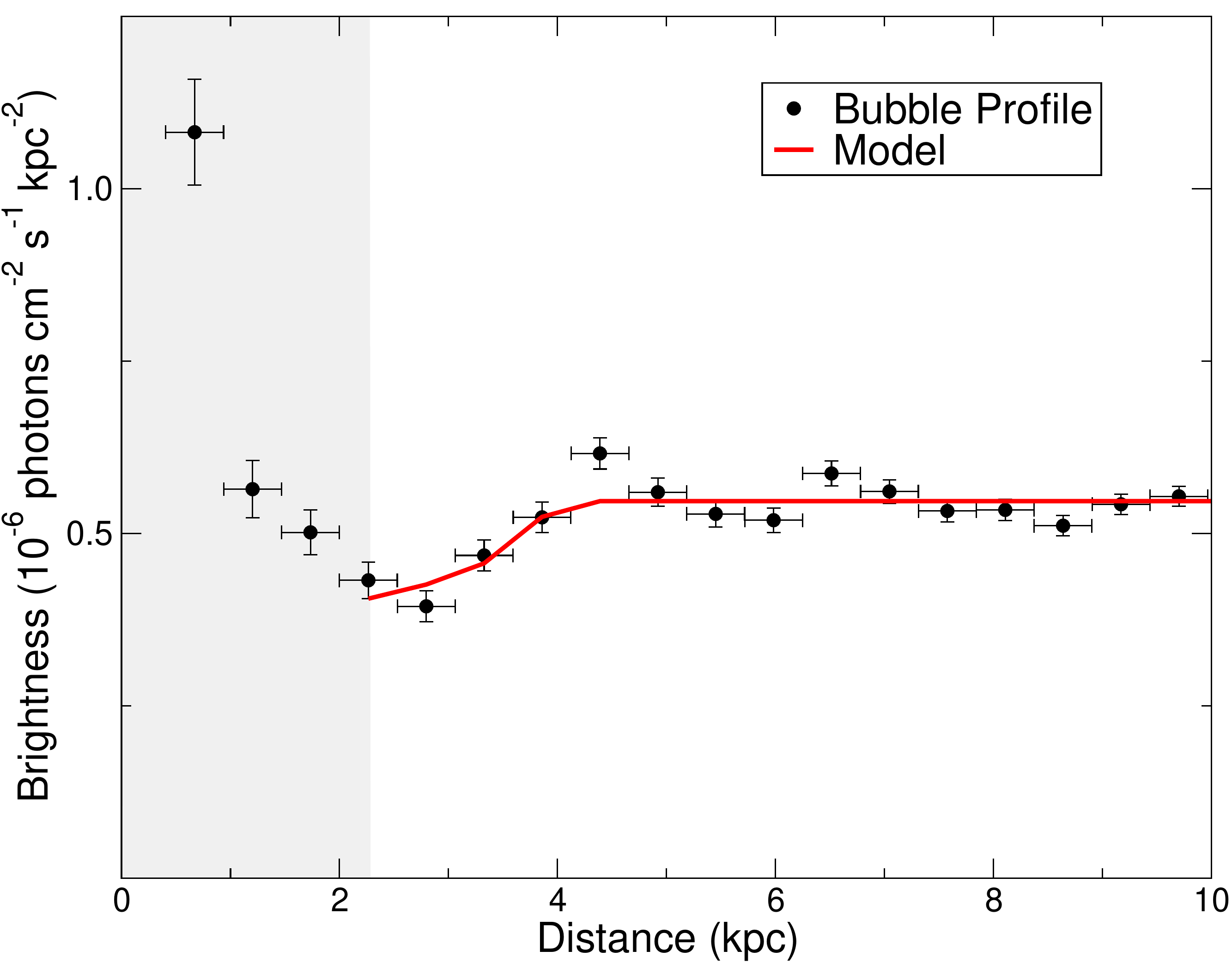}
  \end{tightcenter}
\caption{(Left) Annular regions used for the surface brightness profile of the hole surrounding hotspot~E. The regions were centered on hotspot~E and were selected to avoid jet emission, hotspot~D, and the edge from the cocoon shock. (Right) The extracted surface brightness profile with the fitted model described in Section~\ref{sect:bubble}. The first three data points of the surface brightness profile (gray region) are due to hotspot~E and were therefore omitted from the model fit region.}
\label{fig:bubble}
\end{figure*}

To measure the surface brightness profile for the hole surrounding hotspot~E, annular sectors centered on the hotspot were defined. The sectors were positioned to avoid emission from the jet, hotspot~D, and the edge from the cocoon shock. Each annular region had a radial width of 0.5\arcsec{}, matching the pixel size of \chandra{} ACIS. See the left panel of Figure~\ref{fig:bubble} for an image of the region. The \ciao{} routine {\tt dmextract} was again used to extract the surface brightness from the exposure-corrected image. The surface brightness profile is plotted in the right panel of Figure~\ref{fig:bubble}. A significant deficit in brightness is detected surrounding hotspot~E. 

We initially assumed that the X-ray hole is due to a spherical cavity centered on the hotspot and attempted to measure the X-ray emission per unit volume from within it. Thus, its surface brightness profile was modeled as a spherical cavity with a constant deficit of emission per unit volume, $\ecavity$, representing the difference between the emission per unit volume of the lobe and cavity. This provides the surface brightness model
\begin{equation}
\begin{split}
S(r) = \left\{
        \begin{array}{ll}
            D - 2 \ecavity \sqrt{\rcavity^2 - r^2}, & \quad r < \rcavity \\
            D, & \quad r > \rcavity 
        \end{array}
    \right.
\end{split}
\end{equation}
where $\rcavity$ is the radius of the cavity, $D$ is the surface brightness of the lobe, and the cavity depth is $2\sqrt{\rcavity^2 - r^2}$.

Fitting the model to the cavity was performed using \sherpa~v1 with the Nelder--Mead method and $\chi^2$ statistics \citep{Fruscione2006}. Data points from hotspot~E, the initial three points of the surface brightness profile, were omitted from the fit. A best-fit was found with the fit statistics $\chi^2 = 31.3$ for 13 degrees of freedom ($\dof$). The poor fit is primarily driven by the brightness enhancement at the edge of the hole, as removal of this data point improves the fit to $\chi^2 / \dof = 19.9/12$. This reduction in $\chi^2$ provides marginal evidence for enhanced emission along the western rim of the hole (Section~\ref{sect:dynamics}). The remaining excess is due to low level substructure in the region outside the hole.

\begin{table}
	\caption{Hotspot~E Brightness Profile Best Fit Parameters}
	\label{table:model2}
	\begin{tightcenter}
	\begin{tabular}{ c c c }
		\hline \hline	
		$\rcavity$ &  $\ecavity$  & $D$ \\
		$[\rm arcsec]$ & [$\rm photons\,cm^{-2}$ & [$\rm photons\,cm^{-2}$  \\ 
		& $\rm arcsec^{-3}\,s^{-1}$] & $\rm arcsec^{-2}\,s^{-1}$]  \\ 
		\hline
		$3.54\substack{+0.10\\-0.04}$ & $2.45\substack{+0.35\\-0.32} \times 10^{-8}$ 
			& $5.47\substack{+0.05\\-0.04} \times 10^{-7}$ \\
		\hline
	\end{tabular}
	\end{tightcenter}
\end{table}

The best-fit model parameters are provided in Table~\ref{table:model2}, and the fit is shown in Figure~\ref{fig:bubble}. The deficit in emission per unit volume within the cavity was determined to be $\ecavity = 2.45\substack{+0.35\\-0.32}  \times 10^{-8} \rm\ photons\ cm^{-2}\ arcsec^{-3}\ s^{-1}$. This is 1.7 times larger than the emission per unit volume determined for the adjacent region of the radio lobe, a physical impossibility under our assumptions. Implications of this difference are discussed in Section~\ref{sect:discuss}. 

\section{Hotspot Intensities}
\label{sect:flux}

Previous analyses of Cyg\,A have shown notable differences between hotspot fluxes in the eastern and western lobes, in both radio and X-ray bands \citep[and references therein]{Stawarz2007}. However, the X-ray flux from hotspot~E was too faint to measure in shallower exposures. We therefore sought to quantify the X-ray fluxes of the hotspots using the deep \chandra{} dataset by extracting and fitting their spectra. Our purpose here is to compare the brightnesses of the primary and secondary hotspots in the two lobes; see de Vries et al. (in prep.) for a more thorough study of the hotspots.

\begin{figure}
  \begin{tightcenter}   
	\includegraphics[width=0.99\linewidth]{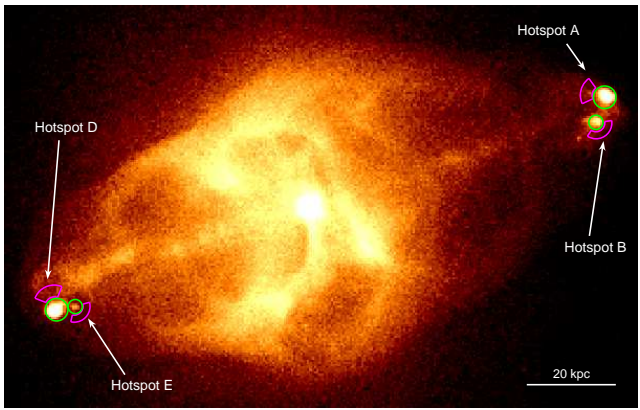}
  \end{tightcenter}
\caption{Regions selected for the hotspot spectral analysis described in Section~\ref{sect:flux}. Green circles mark the source regions, while the adjacent magenta annular regions are used for background. All regions were selected to avoid emission from the jets and other internal structures present within the lobes.} 
\label{fig:hotspots}
\end{figure}

To begin, a region was defined surrounding each hotspot, and a local region within the lobe was defined to represent background emission. The regions were selected to avoid obvious structures within the radio cocoon and sized to account for broadening of the \chandra\ PSF at higher energies. They are marked in Figure~\ref{fig:hotspots}. Spectra were extracted with the {\tt specextract} task in \ciao{} and grouped to obtain a minimum of 80 counts per bin for hotspots A and D, while a minimum of 40 counts per bin was used for hotspots B and E due to their lower fluxes. Each spectrum was fitted with the model ${\tt phabs}({\tt cflux} \times {\tt zpowerlw})$ in \xspec{} v12.10.1b \citep{Arnaud1996}, over the 0.5--7.0 \,keV energy band using $\chi^2$ statistics. The 0.5--7.0\,keV flux and the photon index were left free for each spectrum. The Galactic column density, $N_{\rm H}$, was also left free, but a common value was used for all the spectra as we assume the value does not vary significantly across Cyg\,A.

Parameters for the best-fit models are provided in Table~\ref{table:hotspots}. The Galactic column density was measured as $N_{\rm H} = 4.1 \pm 0.1 \times 10^{21} \rm\ cm^{-2}$. This value is elevated compared to previous estimates \citep{Wright2004, Stawarz2007, Snios2018b}, but the higher value improves the fit noticeably for each hotspot. Power-law indices measured for the hotspots are steeper than previous measurements of Cyg\,A \citep{Wright2004}, though this difference is attributable to the different $N_{\rm H}$ values used. Overall, previous trends for power-law slopes and X-ray fluxes between the hotspots are consistent with our results. We note that, while secondary hotspot~D in the east is appreciably brighter than secondary hotspot~A in the west, the situation is reversed for the primary hotspots, with hotspot~B in the west being $\simeq 3.3$ times brighter than hotspot~E in the east.

\section{Discussion}
\label{sect:discuss}

\subsection{Comparison of Cavity Structure to Radio Observations}
\label{sect:radio}

The primary hotspots are believed to be sites where the jets have most recently encountered the shock compressed ICM that envelopes the radio lobes \citep{Stawarz2007}. The probable cause of the hole in the X-ray emission around hotspot~E is that shocked jet plasma flowing out of the hotspot has displaced the radio lobe plasma in a region around it, creating a cavity that we see as the hole in X-ray brightness. In that case, we might expect to see a related feature at radio wavelengths.

\begin{figure*}
  \begin{tightcenter}   
	\includegraphics[width=0.99\linewidth]{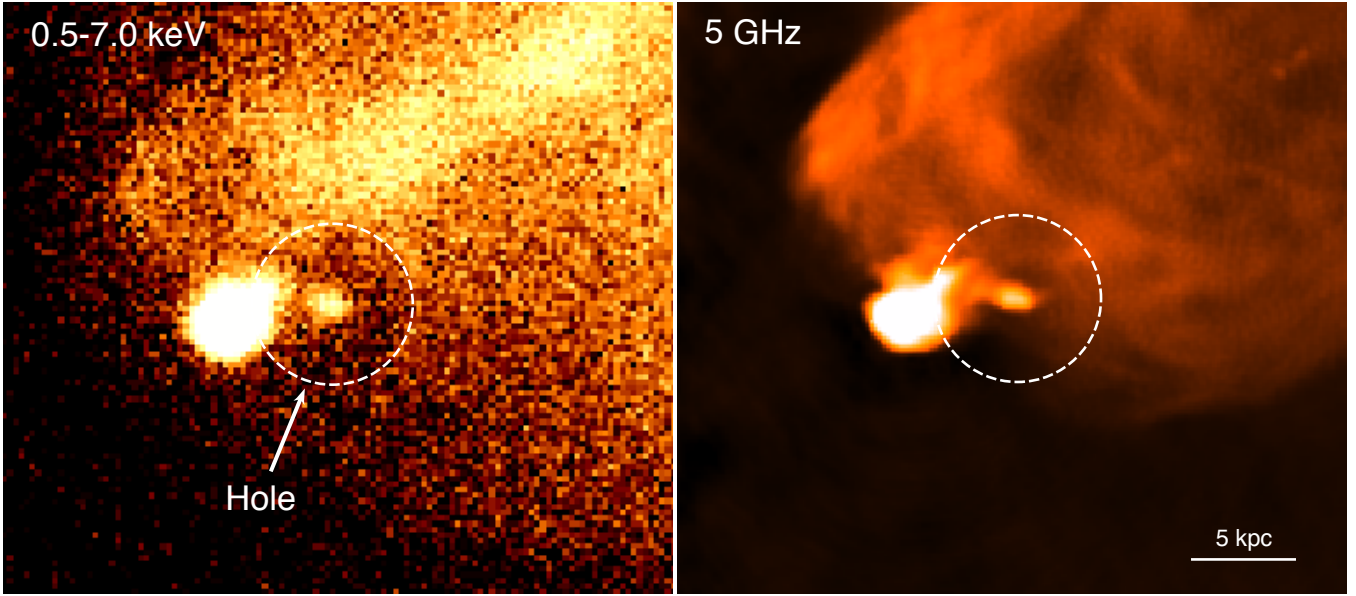}
  \end{tightcenter}
\caption{Comparison between a 0.5--7.0 keV \chandra{} image and a 5\,GHz VLA radio map. The hole is detected in both X-rays and radio, indicating that a cavity has been carved in the radio lobe by outflow from hotspot~E. A comparison of the surface brightness profiles from the two energy bands is provided in Figure~\ref{fig:cavity_comparison}.}
\label{fig:radio}
\end{figure*}

\begin{figure}
  \begin{tightcenter}   
	\includegraphics[width=1.0\linewidth]{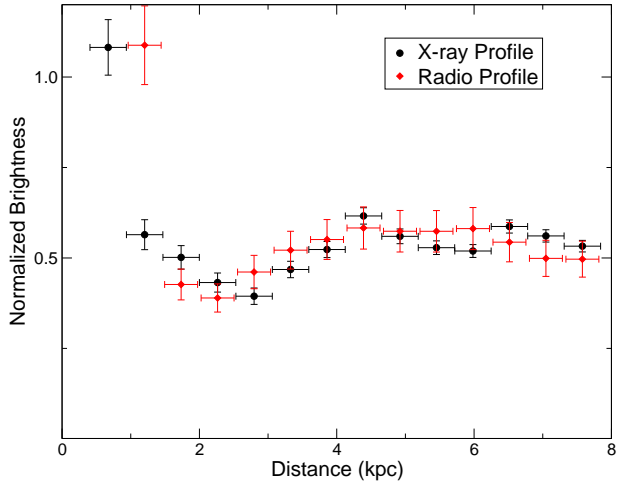}
  \end{tightcenter}
\caption{Comparison between surface brightness profiles from the 0.5--7.0 keV \chandra{} image and a 5\,GHz VLA radio map \citep{Perley1984}. The region used for both cuts is identical to that shown in Figure~\ref{fig:bubble}, left, and the profiles are normalized with respect to the average emission beyond the hole ($>$ 4\,kpc). The feature possess a similar radius and depth in both the X-ray and radio emissions.}
\label{fig:cavity_comparison}
\end{figure}

To investigate the presence of a deficit around hotspot~E in radio, the \chandra{} observations were compared against a 5\,GHz radio map of Cyg\,A observed with the Very Large Array \citep[VLA;][]{Perley1984}. A side-by-side comparison of the datasets is shown in Figure~\ref{fig:radio}, where the dashed circle corresponds to the cavity radius determined for the best fitting model in Section~\ref{sect:bubble}. A deficit in brightness surrounding hotspot~E can be seen in the radio map, so a surface brightness profile was extracted. The annular sectors used to make the X-ray surface brightness profile (Section~\ref{sect:bubble}) were again utilized, where the annular sectors were centered on hotspot~E in the radio map. The measured flux error for the radio maps was estimated by summing in quadrature the background RMS with a 10\% flux uncertainty. The radio surface brightness profile in Figure~\ref{fig:cavity_comparison} shows a deficit surrounding hotspot~E with a similar form to that observed in X-rays. A similar deficit in surface brightness is also seen in new, deep Jansky Very Large Array (JVLA) maps at 5 and 8\,GHz (Sebokolodi et al., in preparation). The agreement in the form of the X-ray and radio features lends weight to the argument that a cavity has been carved in the radio lobe by outflow from hotspot~E. The remainder of our discussion is based on this premise.

\begin{table}
	\caption{Hotspot Fluxes}
	\label{table:hotspots}
	\begin{tightcenter}
	\begin{tabular}{ c | c c }
		\hline \hline
		Hotspot & 0.5--7.0 keV Flux & Photon Index \\ 
		& $[10^{-14}$ erg\,cm$^{-2}$\,s$^{-1}$] & \\ 
		\hline
		A  & $17.6\substack{+0.3\\-0.3}$ & $1.96 \pm 0.03$\\ 
		B  & $3.7\substack{+0.1\\-0.1}$ & $1.86 \pm 0.08$\\ 
		D  & $25.7\substack{+0.4\\-0.3}$ & $1.83 \pm 0.03$\\ 
		E  & $1.1\substack{+0.1\\-0.2}$ & $1.81 \pm 0.20$ \\ 
		\hline
	\end{tabular}
	\end{tightcenter}
	Hotspots fit with {\tt phabs(cflux\,$\times$\,zpowerlw)} model, where $z=0.0561$. Galactic column densities, $N_{\rm H}$, were tied together for all fits and measured as $4.1\pm0.1 \times 10^{21}\rm\ cm^{-2}$.
\end{table}

Properties of the cavity are constrained by the appearance of a hole in both the X-ray and radio. Outside of the X-ray jet, the X-ray emission from the interior of the lobe is known to contribute significantly to the total surface brightness \citep[Figure~\ref{fig:cut}, right;][]{Snios2018b}. Given that the radio emission also originates from the lobe, the presence of a hole in both energy bands lends strong support to the presumption that the deficits are due to a cavity excavated from the western radio lobe of Cyg\,A. The extent of the cavity along our line of sight is estimated in the next section. Examination of the X-ray surface brightness profile of the hole also shows a brightness enhancement along the western edge that is not present in the radio data, shown in Figure~\ref{fig:cavity_comparison}. This feature is discussed in Section~\ref{sect:dynamics}. 

\subsection{Dimensions of the Cavity}
\label{sect:elongation}

Under the assumption that the hole in the X-ray image is due to a spherical cavity, we found in Section~\ref{sect:bubble} that the deficit of X-ray emission per unit volume inside it would need to be $\ecavity = 2.45\substack{+0.35\\-0.32} \times 10^{-8} \rm\ photons\ cm^{-2}\ arcsec^{-3}\ s^{-1}$, exceeding the X-ray emission per unit volume from the lobe, which is not physically possible for an optically thin cavity. The clear implication is that the extent of the cavity on our line of sight exceeds its diameter on the sky. We first consider the possibility that the hole reflects a cavity in the shocked ICM, rather than the radio lobe. To produce the observed X-ray deficit in that scenario, the extent of the cavity along our line of sight would need to be at least a factor of $\ecavity/\eshock$ times the projected diameter of the hole, where $\eshock$ is the emission per unit volume of the shocked ICM. This depth is minimized if the inclination of the radio axis to our line of sight is $\theta=90^\circ$, in which case $\eshock = 6.82\substack{+0.24\\-0.33} \times 10^{-9} \rm\ photons\ cm^{-2}\ arcsec^{-3}\ s^{-1}$, giving a minimum cavity depth of 23\,kpc. Such a depth is well in excess of the observed width for the shocked ICM layer \citep[Figure~\ref{fig:cyga};][]{Snios2018b}. To account for the observed depth, the cavity would need to extend entirely through the shocked ICM layer and tens of kpc further along our line of sight. In addition to failing to account for the deficit in the radio emission, such a cavity is highly implausible.

To account for the deficit in X-ray surface brightness over the hole (Section~\ref{sect:lobe}), it is much more plausible that the cavity has been excavated from within the radio lobe, where the X-ray emission per unit volume peaks on lines of sight through the hole. In that case, the ratio of the cavity depth to its projected diameter would need to be, at least, $\ecavity$/$\elobe$, which is again minimized when the inclination of the radio axis to our line of sight is $\theta=90^\circ$, maximizing $\elobe = 1.42\substack{+0.07\\-0.06} \times 10^{-8} \rm\ photons\ cm^{-2}\ arcsec^{-3}\ s^{-1}$. Thus $\ecavity / \elobe \ge 1.7\pm 0.3$, giving a minimum cavity depth of $13.3\pm2.3$\,kpc. The cavity would need to be deeper for inclinations smaller than $90^\circ$, reducing the emission per unit volume from the lobe (see Appendix~\ref{apen:fit}). Additionally, the depth estimate assumes there is no X-ray emission from the plasma within the cavity. As discussed in Section~\ref{sect:dynamics}, there may well be some emission from this plasma, in which case the cavity would need to be even deeper to cause the same surface brightness deficit. 

\subsection{Jet Dynamics}
\label{sect:dynamics}

The jet entering hotspot~E is thought to be highly supersonic \citep{Krichbaum1998, Boccardi2016, Snios2018b}. When it encounters the compressed ICM at the hotspot, it will pass through one or more shocks, converting jet kinetic energy into thermal and nonthermal particle energy as well as magnetic fields \citep[e.g.,][]{Marcowith2016}. The increased pressure of the shocked jet plasma drives outflow from the hotspot. 

To create a cavity that is deeper than its diameter requires the outflow to be directed roughly along our line of sight. Since it is unlikely that our viewing direction is exactly parallel to the axis of the cavity, its true depth may well be appreciably greater than 1.7 times its diameter. A flow out of hotspot~E that can create such an elongated cavity must be jet-like, suggesting that the jet continues to flow a significant distance beyond hotspot~E. 

Advancing at the speed of light, the time required for the jet to bore a cavity 13.3\,kpc in length into the lobe would be $\simeq$ 40,000\,yr. This is the minimum time required to create the observed cavity, requiring the direction of the deflected jet to remain stable on at least this timescale. We note that the cavity creation time is substantially longer than the time required for a hotspot to dissipate if the jet moves away \citep{Steenbrugge2008, Pyrzas2015}, so hotspots D and E must have both interacted with the jet during the formation period of the cavity. The simultaneous presence of the cavity and hotspots may indicate a causal relationship between them, with the jet being deflected from E to D. A similar scenario where the deflected jet terminates in hotspot~D has been proposed in the past \citep{Williams1985, Cox1991}. 
 
Although its statistical significance is marginal, the enhanced emission on the western rim of the hole discussed in Section~\ref{sect:bubble} may be due to a shock at the interface between the outflowing jet and the lobe plasma. If the shock accelerates electrons to energies high enough to produce X-ray synchrotron emission, the emitting region would naturally be narrow due to the short synchrotron lifetimes. 

The shock at hotspot~E irreversibly converts jet kinetic energy into internal energy, causing the outflowing jet to expand until its pressure matches that of the lobe. These effects will cause the outflowing jet to slow and broaden. In the 43\,GHz radio map \citep{Carilli1999}, the diameter of the jet flowing into hotspot~E appears to be no more than $\simeq$\,0.5\arcsec, whereas the diameter of the outflow determined in Section~\ref{sect:bubble} is closer to $\simeq$\,7\arcsec. Assuming that little of the power carried by the jet is lost in hotspot~E, the power carried as internal energy will have increased substantially after the hotspot. 

\subsection{Doppler Beaming}
\label{sect:beaming}

\begin{figure}
  \begin{tightcenter}   
	\includegraphics[width=0.98\linewidth]{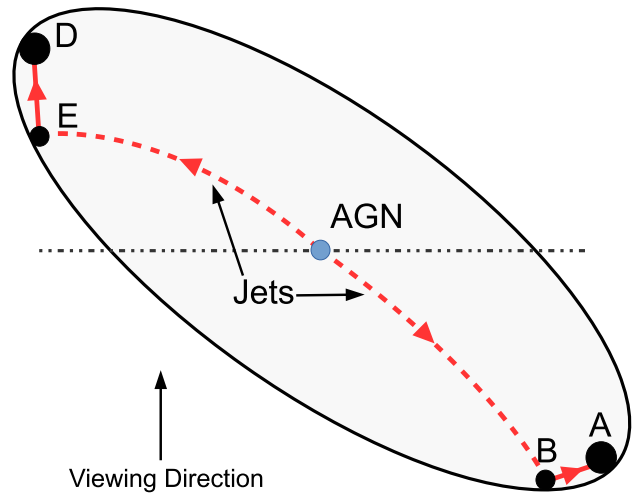}
  \end{tightcenter}
\caption{A schematic, plan diagram of the proposed collision geometry for the jets of Cygnus\,A. The dashed line segments for the jet indicate a causal connection between the jet features rather than the paths of the jets. In order for the outflowing jet from hotspot~E to create a cavity elongated close to our line of sight, the hotspot must be on the near side of the lobe and then the outflow from E is directed away from us. For the figure, we assume that it terminates in hotspot~D.} 
\label{fig:cavity}
\end{figure}

As discussed in Section~\ref{sect:flux}, we found that the 0.5--7.0 keV intensity from hotspot~B, the primary hotspot in the western lobe, is $\simeq 3$ times larger than that from hotspot~E, while hotspot~A in the west is significantly fainter than hotspot~D. The light travel delay between the eastern and western hotspots alone is likely to exceed $10^5$ yr, so we should not expect detailed symmetry between their properties in the two lobes. Even so, the relative faintness of hotspot~E and, particularly, the associated cavity may be explained, at least in part, by the combined effects of the orientation of the radio source and Doppler beaming.

A schematic diagram of the proposed Cyg\,A jet geometry is shown in Figure~\ref{fig:cavity}. Although it is generally accepted that the western jet is approaching, the inclination of the jets is not well determined, with estimates ranging from $55^\circ$ to almost $90^\circ$ \citep{Vestergaard1993, Boccardi2016}. The dashed line segments in the figure indicate the propagation of the jets from the AGN to the primary hotspots, though not their actual paths. If hotspot~E resides on the near side of the lobe, with this orientation the path of the jet from E to D can be directed away from us, approximately along our line of sight. Doppler beaming may then dim any emission from the shocked jet plasma flowing out of the hotspot, which we assume fills the cavity. 

For hotspot~B, \cite{Carilli1988} found an abrupt change in rotation measure that forms an arc around the hotspot. They argue that this feature is due to the shock driven into the ICM by the impact of the jet at the hotspot, making a strong case that hotspot~B also lies on the near side of the lobe. Hotspot~B is also projected to the south of A, near the edge of the lobe, in contrast to hotspot~E, which lies almost due east of hotspot~D (Figure~\ref{fig:cyga}). If the jet continues on from B to A, as indicated in the sketch, also directed significantly northward, the axis of the flow may be more transverse to our line of sight. In this orientation, Doppler beaming would cause significantly less dimming of the outflow than in the east. The depth of the cavity on our lines of sight would also be smaller, helping to explain why no cavity is seen around hotspot~B.

Relative to stationary plasma, Doppler beaming changes the emission from the moving jet plasma by a factor of $D^{2-\alpha}$ \citep{Lind1985}, where the Doppler factor is $D = \sqrt{1 - \beta^2}/(1 - \beta\cos\thetajet)$, $\thetajet$ is the angle between the jet flow and our line of sight ($\thetajet = 0$ for approaching flow), $\beta c$ is the flow speed, and $\alpha$ is the spectral index of the emission. For example, with a flow speed of $\beta = 0.8$ \citep{Snios2018b} and a spectral index of $\alpha = -0.7$ \citep{deVries2018}, relativistic beaming would reduce the emission per unit volume from jet plasma flowing directly away from us by a factor $\simeq 0.05$. This is clearly sufficient to create the appearance of a hole. For flow perpendicular to our line of sight the reduction factor would be $\simeq 0.25$, still sufficient to create an apparent hole. However, as noted above, the depth of the cavity on our line of sight would also be decreased, significantly reducing its contrast with respect to the lobe. 

Doppler beaming may additionally account for the observed factor of 3 difference in flux from the primary hotspots B and E. Using a flow speed of $\beta = 0.8$ and the spectral indices from Table~\ref{table:hotspots}, Doppler beaming will produce a flux difference $\ge 3$ between the hotspots in cases where the deflected jet angle of hotspot~E relative to our line of sight $\theta_{\rm jet, E}$ is less than the hotspot~B jet angle $\theta_{\rm jet, B}$, or $\theta_{\rm jet, E} < \theta_{\rm jet, B}$. This parameter range is broadly consistent with the angles inferred from observations, indicating that Doppler beaming can plausibly explain the difference in intensity of the primary hotspots. Future analysis of the three-dimensional structure in Cyg\,A via radio and X-ray observations will further constrain this parameter range and consequently test the validity of the proposed Doppler beaming scenario.  

\section{Conclusions}
\label{sect:conclusions}

\chandra{} observations of Cygnus\,A, totaling 2\,Msec, have revealed an approximately circular hole, with a radius of 3.9\,kpc, centered on primary hotspot~E in the eastern radio lobe. X-ray surface brightness profiles were extracted and fitted for the hole and the surrounding radio lobe, allowing us to infer the X-ray emission per unit volume for each region. Based on these results, it is far more likely that the hole is due to a cavity excavated from the lobe than any other region. Assuming this, the surface brightness profile of the hole requires a minimum depth $1.7\pm0.3$ times greater than its projected width. The cavity depth is therefore at least $13.3\pm2.3$\,kpc, significantly elongated on our line of sight. 

To investigate the presence of the hole in other energy bands, the co-added X-ray observation was compared with an archival 5\,GHz VLA radio map of Cygnus\,A. A hole surrounding hotspot~E was also observed in the radio, with a radius approximately equal to that measured from X-rays. A radio surface brightness profile from the hole was extracted with the same annular sectors used in the X-ray analysis, and the radio profile is similar in form to the X-ray profile, suggesting that there is a hole of similar dimensions in the radio lobe. This agreement between the radio and X-ray reinforces the argument that outflow from hotspot~E has carved an elongated cavity in the radio lobe.

X-ray hotspot fluxes were measured for Cygnus\,A, placing the best constraints on both the hotspot fluxes and the Galactic column density $N_{\rm H}$ for the system, to date. 
Hotspot~E in the east was found to be $\simeq 3$ times dimmer in X-rays than the corresponding primary hotspot~B in the west. To explain this discrepancy, we argue that the orientation of Cygnus A causes the outflow at hotspot~E to travel almost directly away from us, creating an elongated cavity. For primary hotspot~B in the west, the outgoing jet may be deflected more across our line of sight. Doppler beaming can readily account for the apparent discrepancy in intensities of the primary hotspots as well as the lack of emission from the cavity associated with hotspot~E. Altogether, these results favor the deflected jet scenario, in which the jet of Cygnus\,A is deflected at a primary hotspot, then it travels onward to deposit the bulk of its power at a secondary hotspot. There may be more than one deflection before the jet ultimately terminates, resulting in multiple active hotspots in a lobe. The results disfavor the dentist drill model. 

\acknowledgements{
Support for this work was provided by the National Aeronautics and Space Administration through \chandra{} Award Number G07-18104X issued by the \chandra\ X-ray Observatory Center, which is operated by the Smithsonian Astrophysical Observatory for and on behalf of the National Aeronautics Space Administration under contract NAS8-03060. P.E.J.N. and R.P.K. were supported in part by NASA contract NAS8-03060. We additionally thank C. Carilli for his invaluable comments on this work. 
}

\software{
\ciao{} v4.10 \citep{Fruscione2006}, \sherpa{} v1 \citep{Freeman2001}, \xspec{} v12.10.1b \citep{Arnaud1996}
}

\bibliographystyle{aasjournal}
\bibliography{all_data}

\appendix 

\section{Lobe Emission Model} 

\subsection{Model}
\label{apen:model}

The cluster is divided into three regions representing the undisturbed ICM, the layer of shock compressed ICM between the cocoon shock and the outer edge of the radio lobe, and the interior of the radio lobe. The X-ray emission per unit volume from the undisturbed ICM is assumed to be well-described by a beta model \citep{Jones1984}. The X-ray emission per unit volume from the shock compressed ICM is modeled as constant, as is the emission from within the radio lobe. The surface brightness profile is calculated by projecting the three-dimensional emission distribution along lines of sight.

The emission per unit volume for a beta model has the form
\begin{equation}
 A (1 + r^2/a^2)^{-3\beta},
\end{equation}
where $r$ is distance from the AGN, which is assumed to lie at the cluster center. It is convenient to use cartesian coordinates, with the origin at the AGN and the $z$ axis parallel to the line of sight. Positions on the sky are defined by their $x$ and $y$ coordinates. To project the beta model onto the sky, we need to evaluate integrals of the form
\begin{equation} \label{eqn:intbeta}
 \int_z^\infty A \left[1 + (x^2 + y^2 + z'^2) / a^2\right]^{-3\beta} dz'
 = \frac{Aa}{2} \left[1 + (x^2 + y^2) / a^2\right]^{1/2 - 3 \beta}
 f (x, y, z),
\end{equation}
where
\begin{equation}
 f (x, y, z) =
 \begin{cases}
  B (3\beta - 0.5, 0.5; t), &z \ge0 , \\
  2 B (3\beta - 0.5, 0.5) - B (3\beta - 0.5, 0.5; t), &z <0,
 \end{cases}
\end{equation}
with $t = (a^2 + x^2 + y^2) / (a^2 + x^2 + y^2 + z^2)$. Here $B(u,v;t)$ is the incomplete beta function and $B(u,v) = B(u, v; 1)$ is the beta function. Note that, since the beta function is even in $z$, the integral of the beta model from $-\infty$ to $z$ is obtained simply by replacing $f(x,y,z)$ in equation (\ref{eqn:intbeta}) by $f(x,y, -z)$. 

The radio cocoon is assumed to be symmetric under rotation about an axis through the AGN, coinciding approximately with the radio jets. The $x$-axis is oriented parallel to the projection onto the sky of this symmetry axis. In terms of these coordinates, the surface brightness profile of the lobe was measured at fixed $x$, so that it is a function of $y$ alone. If the inclination, $\theta$, of the cocoon axis to our line of sight is not much less than $90^\circ$, the segment of the cocoon that contributes to the surface brightness profile is small. This permits us to approximate the cocoon shock and outer edge of the lobe in the region of interest as a pair of nested cylinders. For the cylinder of radius $\rhoc$ representing the cocoon shock, the line of sight at projected position $(x, y)$, with $|y| < \rhoc$, will intersect the cylinder at the two positions
\begin{equation}
 \zcboth = \left(x \cos\theta \pm \sqrt{\rhoc^2 - y^2}\right) / \sin\theta ,
\end{equation}
(we assume $\sin\theta > 0$ with no loss of generality). The length of this line segment within the cylinder is 
\begin{equation}
 \zcp - \zcm = 2 \sqrt{\rhoc^2 - y^2} / \sin\theta.
\end{equation}
Similarly, if the radius of the cylinder representing the radio lobe is $\rhol$, for $|y| < \rhol$, the length of the line of sight within the lobe will be
\begin{equation}
 2 \sqrt{\rhol^2 - y^2} / \sin\theta.
\end{equation}
Denoting the X-ray emission per unit volume within the lobe by $\elobe$ and that within the shock compressed ICM by $\eshock$, the form of the surface brightness profile for sight lines that intersect the lobe, \ie, $|y| < \rhol$, would then be
\begin{equation}
 \begin{split}
  S(y) = \frac{Aa}{2} &\left[1 + (x^2 + y^2) / a^2\right]^{1/2 - 3 \beta}
   \left[f(x, y, -\zcm) + f(x, y, \zcp)\right] \\ 
   &+ \frac{2}{\sin\theta} \left[(\elobe - \eshock) \sqrt{\rhol^2 - 
     y^2} + \eshock \sqrt{\rhoc^2 - y^2} \right] + C,
 \end{split}
\end{equation}
for sight lines that miss the lobe, but pass through the shock compressed ICM, \ie, $\rhol \le |y| < \rhoc$, it would be
\begin{equation}
 S(y) = \frac{Aa}{2} \left[1 + (x^2 + y^2) / a^2\right]^{1/2 - 3 \beta}
 \left[f(x, y, -\zcm) + f(x, y, \zcp)\right] + \frac{2}{\sin\theta}
 \eshock \sqrt{\rhoc^2 - y^2} + C,
\end{equation}
and for sight lines that miss the cocoon completely, for \ie, $|y| \ge
\rhoc$, it would be
\begin{equation}
 S(y) = \frac{Aa}{2} \left[1 + (x^2 + y^2) / a^2\right]^{1/2 - 3 \beta}
 f(x, y, -\infty) + C.
\end{equation}
In all cases, a constant background emission $C$ is added. 

\subsection{Model Fitting Method and Best-Fit Parameters}
\label{apen:fit}

The lobe surface brightness model was fit to the data using \sherpa{} v1 with the Nelder--Mead method and $\chi^2$ statistics \citep{Fruscione2006}. The X-ray jet is brighter than the remainder of the lobe \citep[de\,Vries et al., in preparation]{Steenbrugge2008}, and was not included in the fit. Additionally, emission from the northern half of the radio lobe was excluded due to the observed asymmetry between the northern and southern sections, with the northern half being consistently brighter. $x$ was fixed at a distance of 45.17\arcsec, making the model a function of $y$ alone. Using an inclination angle $\theta$ of 55$^{\circ}$ \citep{Vestergaard1993}, a best-fit was found with the fit statistics \mbox{$\chi^2/\dof = 84.6/66$.} The best-fit model is overlaid on the brightness profile in Figure~\ref{fig:cut}. 

Inclination angle was varied between $ 20^{\circ} < \theta < 90^{\circ}$ to investigate its impact on each model parameter. The best-fit model parameters for $\theta = 55^\circ$ are provided in Table~\ref{table:model1}, and the best-fit for $\theta = 90^\circ$ are provided in Table~\ref{table:model3}. $\beta$, $a$, and $A$ were shown to increase with decreasing $\theta$. $\rhoc$, $\rhol$, and $C$ were insensitive to changes in $\theta$. The best-fit results for $\eshock$ and $\elobe$ were found to decrease with decreasing $\theta$, but the overall changes were $< 3\sigma$. We therefore utilized the results from the $\theta = 55^{\circ}$ model for this work. 

\begin{table}[h]
	\caption{Radio Lobe Brightness Profile Best-fit Parameters for $\theta = 55^\circ$}
	\label{table:model1}
	\begin{tightcenter}
	\begin{tabular}{ c c c c c c c c }
		\hline \hline
		$\beta$ & $a$ & $\rhoc$ & $\rhol$ & $A$ 
			& $\eshock$ & $\elobe$ & $C$ \\ 
		& [arcsec] & [arcsec] & [arcsec] & [$\rm photons\,cm^{-2}$ 
			& [$\rm photons\,cm^{-2}$ & [$\rm photons\,cm^{-2}$ 
			& [$\rm photons\,cm^{-2}$  \\ 
		& & & & $\rm arcsec^{-3}\,s^{-1}$] & $\rm arcsec^{-3}\,s^{-1}$] 
			& $\rm arcsec^{-3}\,s^{-1}$] & $\rm arcsec^{-2}\,s^{-1}$] \\ 
		\hline
		$0.58\substack{+0.01\\-0.01}$ & $0.14\substack{+0.01\\-0.01}$ 
			& $19.8\substack{+0.3\\-0.3}$ & $9.4\substack{+0.2\\-0.2}$  
			& $2.28\substack{+0.05\\-0.04}$ 
			& $5.73\substack{+0.22\\-0.18} \times 10^{-9}$ 
			& $1.11\substack{+0.05\\-0.06} \times 10^{-8}$ 
			& $1.74\substack{+0.59\\-0.38} \times 10^{-8}$ \\
		\hline
	\end{tabular}
	\end{tightcenter}
\end{table}

\begin{table}[h]
	\caption{Radio Lobe Brightness Profile Best-fit Parameters for $\theta = 90^\circ$}
	\label{table:model3}
	\begin{tightcenter}
	\begin{tabular}{ c c c c c c c c }
		\hline \hline
		$\beta$ & $a$ & $\rhoc$ & $\rhol$ & $A$ 
			& $\eshock$ & $\elobe$ & $C$ \\ 
		& [arcsec] & [arcsec] & [arcsec] & [$\rm photons\,cm^{-2}$ 
			& [$\rm photons\,cm^{-2}$ & [$\rm photons\,cm^{-2}$ 
			& [$\rm photons\,cm^{-2}$  \\ 
		& & & & $\rm arcsec^{-3}\,s^{-1}$] & $\rm arcsec^{-3}\,s^{-1}$] 
			& $\rm arcsec^{-3}\,s^{-1}$] & $\rm arcsec^{-2}\,s^{-1}$] \\ 
		\hline
		$0.54\substack{+0.01\\-0.01}$ & $0.07\substack{+0.01\\-0.01}$ 
			& $19.8\substack{+0.4\\-0.2}$ & $9.4\substack{+0.2\\-0.2}$ 
			& $3.24\substack{+1.23\\-0.03}$ 
			& $6.82\substack{+0.24\\-0.33} \times 10^{-9}$ 
			& $1.42\substack{+0.07\\-0.06} \times 10^{-8}$ 
			& $1.96\substack{+0.53\\-0.39} \times 10^{-8}$ \\
		\hline
	\end{tabular}
	\end{tightcenter}
\end{table}

\end{document}